\documentstyle{article}

\newcommand{\h}{\hspace{.5 em}}
\newcommand{\bdm}{\begin{displaymath}}
\newcommand{\edm}{\end{displaymath}}
\newcommand{\bi}{\begin{itemize}}
\newcommand{\ei}{\end{itemize}}
\newcommand{\be}{\begin{equation}}
\newcommand{\ee}{\end{equation}}
\newcommand{\bea}{\begin{eqnarray}}
\newcommand{\eea}{\end{eqnarray}}
\newcommand{\beas}{\begin{eqnarray*}}
\newcommand{\eeas}{\end{eqnarray*}}
\begin{document}
\title{Deformation Quantization of sdiff($\Sigma_{2}$) SDYM Equation}
\author{
 M.Przanowski$^{1, 2,}$\footnote{e-mail:przan@fis.cinvestav.mx},\and
J.F.Pleba\'{n}ski$^{1,}$\footnote{e-mail:pleban@fis.cinvestav.mx}
\and and S.Forma\'{n}ski$^{2,}$\footnote{e-mail:sforman@ck-sg.p.lodz.pl} } 
\maketitle
\noindent $^{1}$Departamento de Fisica, CINVESTAV, 07000 Mexico D.F., Mexico
\vspace{2 ex}

\noindent $^{2}$Institute of Physics, Technical University of \L\'{o}d\'{z}, 
W\'{o}lcza\'{n}ska 219. 93-005 \L\'{o}d\'{z}, Poland

\vspace {2 ex}

\noindent\bf ABSTRACT \rm  Deformation quantization (the Moyal deformation) of SDYM equation 
for the algebra of the area preserving  diffeomorphisms of a 2-surface $\Sigma_{2}$, 
sdiff($\Sigma_{2}$), is studied. Deformed equation we call the \it master equation \rm
(ME) as it can be reduced to many integrable nonlinear equations in mathematical physics.
Two sets of concerved charges for ME are found. Then the linear systems for ME (the Lax pairs) 
associated with the conserved charges are given. We obtain the dressing operators and the 
infinite algebra of hidden symmetries of ME. Twistor construction is also done.

%
%
\section{Introduction}
In 1994 V.Husain \cite{1} was able to reduce the Ashtekar-Jacobson-Smolin equations 
describing the metric of self-dual complex vacuum spacetimes (the heavenly spacetimes) 
to one equation for one holomorphic function $\Theta_{0}=\Theta_{0}(x,y,p,q)$
\be
\label{H-P_eqn}
\partial^{2}_{x}\Theta_{0}+\partial^{2}_{y}\Theta_{0}+\{\partial_{x}\Theta_{0},
\partial_{y}\Theta_{0}\}_{{\cal P}}=0
\ee
where $\{\cdot , \cdot\}_{{\cal P}}$ stands for the Poisson bracket
\be
\{f , g \}_{{\cal P}} = f \stackrel{\leftrightarrow}{{\cal P}} g\, , \h\h\h
\stackrel{\leftrightarrow}{{\cal P}}:=\frac{\stackrel{\leftarrow}{\partial}}
{\partial q}\frac{\stackrel{\rightarrow}{\partial}}
{\partial p}-\frac{\stackrel{\leftarrow}{\partial}}{\partial p}
\frac{\stackrel{\rightarrow}{\partial}}{\partial q}
\ee
Eq. (\ref{H-P_eqn}) is called the \it Husain-Park heavenly equation \rm (\it 
H-P equation\rm ) as it has been also found by Q.H.Park \cite{2} from another
point of view. Namely, in Park's approach Eq. (\ref{H-P_eqn}) is the principal
chiral model equation for the algebra of the area preserving diffeomorphisms 
of a 2-surface $\Sigma_{2}$, sdiff($\Sigma_{2}$), and it is obtained by a symmetry
reduction of sdiff($\Sigma_{2}$) SDYM equation
\be
\label{Park}
\partial_{x}\partial_{\tilde{x}}\Theta_{0}+\partial_{y}\partial_{\tilde{y}}\Theta_{0}+
\{\partial_{x}\Theta_{0},\partial_{y}\Theta_{0}\}_{{\cal P}}=0
\ee
where now $\Theta_{0}=\Theta_{0}(x,y,\tilde{x},\tilde{y},p,q)$.

\noindent A natural generalization of Eq. (\ref{Park}) can be done when  the Poisson 
bracket is changed by the Moyal one. Thus one arrives at the following equation \cite{3,4}
\bea
\label{ME}
\partial_{x}\partial_{\tilde{x}}\Theta+\partial_{y}\partial_{\tilde{y}}\Theta+
\{\partial_{x}\Theta,\partial_{y}\Theta\}_{{\cal M}}&=&0\\
\Theta=\Theta(\hbar;x,y,\tilde{x},\tilde{y},p,q)& &\nonumber
\eea
where $ \{ \cdot , \cdot \}_{{\cal M}} $ denotes the Moyal bracket 
\begin{eqnarray}
\label{1.8}
& \{ f , g \}_{{\cal M}} :=  \frac{1}{i \hbar} (f \ast g - g \ast f) 
  =  f\frac{2}{ \hbar} sin(\frac{ \hbar}{2} \stackrel{\leftrightarrow}{{\cal P}})g ;
\hspace{1 em}\hbar\epsilon \mbox{\boldmath $R$} & \nonumber\\
& f \ast g :=  \sum_{n=0}^{\infty} \frac{1}{n!} (\frac{i\hbar}{2})^{n}
\omega^{i_1 j_1}\dots\omega^{i_n j_n}
\frac{\partial^{n}f}{\partial X^{i_1}\dots \partial X^{i_n}}
\frac{\partial^{n}g}{\partial X^{j_1}\dots \partial X^{j_n}} &\nonumber\\
 & =  f \exp(\frac{i \hbar}{2} \stackrel{\leftrightarrow}{{\cal P}})g,
\hspace{1 em} i_1 ,\dots j_1 , \dots=1,2; &\nonumber \\  
& (X^1, X^2)=(q,p),\h\h (\omega^{ij})=\left (\begin{array}{cc}
0&1\\
-1&0
\end{array}
\right )&
\end{eqnarray}
The real parameter $\hbar$ is a \it deformation parameter\rm .

\noindent Eq. (\ref{ME}) we call the \it master equation \rm (ME). By a symmetry
reduction, using also some representations of the Moyal bracket Lie algebra one can
reduce ME to the known heavenly equations, to $su(N)$ SDYM equations or $su(N)$
principal chiral model equations and to many integrable nonlinear equations of 
mathematical physics \cite{5}. We should point out that $\Theta(\hbar;x,y,\tilde{x},\tilde{y},p,q)$ being a solution of ME is considered
to be the formal series with respect to $\hbar$,
\be
\label{formal-expansion}
\Theta=\sum_{n=-N}^{\infty}\Theta_{n}\hbar^{n}\h ,\h\h
N<\infty\h ,\h\h \Theta_{n}=\Theta_{n}(x,y,\tilde{x},\tilde{y},p,q)
\ee
In our recent paper \cite{6} some evidence for the integrability of ME has been 
provided. It has been shown that ME admits infinite number of nonlocal conservation
laws and linear systems (Lax pairs) for ME have been found. Moreover, a twistor 
construction has been also done.

The aim of the present work is to consider in some details and develope the results
of \cite{6}. First, in Section 2, we find infinite number of new nonlocal conservation
laws such that the conserved "charges" define hidden symmetries of ME.
In Section 3 new linear systems for ME are obtained and the dressing operators
leading to solutions of these systems are found. The dressing operators appear to be
the solutions of the linear systems presented in \cite{6}.
In Section 4 the infinite Lie algebra of the hidden symmetries of ME is given.
Finally, Section 5 is devoted to a twistor construction for ME.
%
%
\section{Conservation laws and hidden symmetries.}
\setcounter{equation}{0}
Let $\eta^{(0)}=\eta^{(0)}(\hbar;x,y,\tilde{x},\tilde{y},p,q)$ be some function. Define
\be
j^{(1)}_{x}:={\cal D}_{\tilde{x}}\eta^{(0)}\h ,\h\h
j^{(1)}_{y}:={\cal D}_{\tilde{y}}\eta^{(0)}
\ee
where
\be
\label{def-of-D}
{\cal D}_{\tilde{x}}:=\partial_{\tilde{x}}-\frac{1}{i\hbar}\partial_{y}\Theta\ast\h ,\h\h
{\cal D}_{\tilde{y}}:=\partial_{\tilde{y}}+\frac{1}{i\hbar}\partial_{x}\Theta\ast
\ee
and $\Theta=\Theta(\hbar;x,y,\tilde{x},\tilde{y},p,q)$ is a solution of ME (\ref{ME}).
We have
\bea
\partial_{x}j^{(1)}_{x}  +  \partial_{y}j^{(1)}_{y}\, =\,
(\partial_{x}{\cal D}_{\tilde{x}}+\partial_{y}{\cal D}_{\tilde{y}})\eta^{(0)}\, =\,
({\cal D}_{\tilde{x}}\partial_{x}+{\cal D}_{\tilde{y}}\partial_{y})\eta^{(0)}\nonumber\\
=\,\partial_{x}\partial_{\tilde{x}}\eta^{(0)}  + 
\partial_{y}\partial_{\tilde{y}}\eta^{(0)}+\frac{1}{i\hbar}
(\partial_{x}\Theta\ast\partial_{y}\eta^{(0)}-
\partial_{y}\Theta\ast\partial_{x}\eta^{(0)})
\eea
Consequently $\partial_{x}j^{(1)}_{x}  +  \partial_{y}j^{(1)}_{y}=0$ iff $\eta^{(0)}$
satisfies the following linear equation
\be
\label{adjointME}
\partial_{x}\partial_{\tilde{x}}\eta^{(0)}  + 
\partial_{y}\partial_{\tilde{y}}\eta^{(0)}+\frac{1}{i\hbar}
(\partial_{x}\Theta\ast\partial_{y}\eta^{(0)}-
\partial_{y}\Theta\ast\partial_{x}\eta^{(0)})=0
\ee
Given a function $\eta^{(0)}$ satisfying Eq.(\ref{adjointME}), there exists a function
\mbox{$\eta^{(1)}=\eta^{(1)}(\hbar;x,y,\tilde{x},\tilde{y},p,q)$} such that
\bea
\label{2.5}
\partial_{x}\eta^{(1)}&=&{\cal D}_{\tilde{y}}\eta^{(0)}\nonumber\\
-\partial_{y}\eta^{(1)}&=&{\cal D}_{\tilde{x}}\eta^{(0)}
\eea
From (\ref{2.5}) with (\ref{def-of-D}) one gets
\bea
({\cal D}_{\tilde{x}}\partial_{x}+{\cal D}_{\tilde{y}}\partial_{y})\eta^{(1)}\, =\,
({\cal D}_{\tilde{x}}{\cal D}_{\tilde{y}}-{\cal D}_{\tilde{y}}{\cal D}_{\tilde{x}})
\eta^{(0)}\, \nonumber\\
= \frac{1}{i\hbar}(
\partial_{x}\partial_{\tilde{x}}\Theta+\partial_{y}\partial_{\tilde{y}}\Theta+
\{\partial_{x}\Theta,\partial_{y}\Theta\}_{{\cal M}}
)\ast\eta^{(0)}\stackrel{\mbox{\footnotesize by ME \normalsize}}
{=}0
\eea
Therefore, as $\Theta$ is a solution of ME $\eta^{(1)}$ satisfies the same equation
(\ref{adjointME}) as $\eta^{(0)}$ does. This enables us to define the current $j^{(2)}$
\be
j^{(2)}_{x}:={\cal D}_{\tilde{x}}\eta^{(1)}\h ,\h\h
j^{(2)}_{y}:={\cal D}_{\tilde{y}}\eta^{(1)}
\ee
which satisfies the equation $\partial_{x}j^{(1)}_{x}  +  \partial_{y}j^{(1)}_{y}=0$.
Hence there exists a function $\eta^{(2)}$ such that
\bea
\partial_{x}\eta^{(2)}&=&{\cal D}_{\tilde{y}}\eta^{(1)}\nonumber\\
-\partial_{y}\eta^{(2)}&=&{\cal D}_{\tilde{x}}\eta^{(1)}
\eea
Continuing this procedure we arrive at the series of functions (\it conserved charges\rm )
$\eta^{(1)},\eta^{(2)},...$ and currents $j^{(1)},j^{(2)},...$ , defined by the recursion
equations
\bea
j^{(n+1)}_{x}& = & -\partial_{y}\eta^{(n+1)}\, =\, {\cal D}_{\tilde{x}}\eta^{(n)}\nonumber\\
j^{(n+1)}_{y}& = & \partial_{x}\eta^{(n+1)}\,\, =\, {\cal D}_{\tilde{y}}\eta^{(n)}\h ,\h\h
n=0,1,...
\eea
It is evident that as $\Theta$ satisfies ME and $\eta^{(0)}$ satisfies (\ref{adjointME})
all $\eta^{(n)}$ satisfy the linear equation
\be
\label{adjoint-n-ME}
\partial_{x}\partial_{\tilde{x}}\eta^{(n)}  + 
\partial_{y}\partial_{\tilde{y}}\eta^{(n)}+\frac{1}{i\hbar}
(\partial_{x}\Theta\ast\partial_{y}\eta^{(n)}-
\partial_{y}\Theta\ast\partial_{x}\eta^{(n)})=0
\ee
Thus we obtain infinite number of conservation laws
\be
\partial_{x}j^{(n)}_{x}  +  \partial_{y}j^{(n)}_{y}=0\h ,\h\h n=1,2,...
\ee
and the conserved charges
\be
\label{eta-integralform}
\eta^{(n)}=\int^{x}dx^{(n)}{\cal D}_{\tilde{y}}\int^{x^{(n)}}dx^{(n-1)}{\cal D}_{\tilde{y}}
...\int^{x^{(2)}}dx^{(1)}{\cal D}_{\tilde{y}}\eta^{(0)}\h ,\h\h n=1,2,...
\ee
In particular, an interesting case is when one puts
\be
\label{eta-zero=1}
\eta^{(0)}=1.
\ee
Then (taking appropriate boundary conditions) we get from (\ref{2.5}) with (\ref{eta-zero=1})
\be
\label{eta-1-theta}
\eta^{(1)}=\frac{1}{i\hbar}\Theta
\ee
Observe that Eq. (\ref{adjoint-n-ME}) for $\eta^{(1)}$ given by (\ref{eta-1-theta}) is
exactly ME. [The case of $\eta^{(0)}=1$ has been analyzed in our previous work \cite{6}
and in fact the considerations of \cite{6} closely follow E.Brezin et al \cite{7},
M.K.Prasad et al \cite{8}, L.L.Chau et al \cite{9} and L.L.Chau \cite{10} where
nonlocal conservation laws for some 2-dimensional nonlinear field theories  and for
SDYM equations have been found.]

Now we are going to look for another collection of conserved charges. Here we follow
V.Husain \cite{1} and M.Dunajski and L.J.Mason \cite{11} who obtained infinite number of
conservation laws for some heavenly equations in four dimensions.\footnote{ We are 
indebted to Maciej Dunajski for pointing out to us the method how to obtain new conservation laws.}

Assume that $\sigma^{(0)}=\sigma^{(0)}(\hbar;x,y,\tilde{x},\tilde{y},p,q)$ is any solution
of the \it linearized master equation \rm (LME)
\be
\label{LME}
\partial_{x}\partial_{\tilde{x}}\sigma^{(0)}+
\partial_{y}\partial_{\tilde{y}}\sigma^{(0)}+
\{\partial_{x}\Theta,\partial_{y}\sigma^{(0)}\}_{{\cal M}}-
\{\partial_{y}\Theta,\partial_{x}\sigma^{(0)}\}_{{\cal M}}=0
\ee
Define
\be
J^{(1)}_{x}:={\cal L}_{\tilde{x}}\sigma^{(0)}\h ,\h\h 
J^{(1)}_{y}:={\cal L}_{\tilde{y}}\sigma^{(0)}
\ee
where
\be
{\cal L}_{\tilde{x}}:=\partial_{\tilde{x}}-\{\partial_{y}\Theta,\cdot\}_{{\cal M}}\h ,\h\h
{\cal L}_{\tilde{y}}:=\partial_{\tilde{y}}+\{\partial_{x}\Theta,\cdot\}_{{\cal M}}.
\ee
Then
\bea
\partial_{x}J^{(1)}_{x}+\partial_{y}J^{(1)}_{y}=
(\partial_{x}{\cal L}_{\tilde{x}}+\partial_{y}{\cal L}_{\tilde{y}})\sigma^{(0)}=
({\cal L}_{\tilde{x}}\partial_{x}+{\cal L}_{\tilde{y}}\partial_{y})\sigma^{(0)}=\nonumber\\
\partial_{x}\partial_{\tilde{x}}\sigma^{(0)}+
\partial_{y}\partial_{\tilde{y}}\sigma^{(0)}+
\{\partial_{x}\Theta,\partial_{y}\sigma^{(0)}\}_{{\cal M}}-
\{\partial_{y}\Theta,\partial_{x}\sigma^{(0)}\}_{{\cal M}}
\stackrel{\mbox{by (\ref{LME})}}{=}0.
\eea
Hence, there exists a function $\sigma^{(1)}$ such that
\be
\label{2.19}
\partial_{x}\sigma^{(1)}={\cal L}_{\tilde{y}}\sigma^{(0)}\h ,\h\h
-\partial_{y}\sigma^{(1)}={\cal L}_{\tilde{x}}\sigma^{(0)}.
\ee
From (\ref{2.19}) we get
\bea
&({\cal L}_{\tilde{x}}\partial_{x}+{\cal L}_{\tilde{y}}\partial_{y})\sigma^{(1)}=
({\cal L}_{\tilde{x}}{\cal L}_{\tilde{y}}-
{\cal L}_{\tilde{y}}{\cal L}_{\tilde{x}})\sigma^{(0)}&\nonumber\\
&=\{\,\partial_{x}\partial_{\tilde{x}}\Theta+\partial_{y}\partial_{\tilde{y}}\Theta+
\{\partial_{x}\Theta,\partial_{y}\Theta\}_{{\cal M}}\, ,\,\sigma^{(0)}\,\}_{{\cal M}}
\stackrel{\mbox{by ME}}{=}0.&
\eea
It means that $\sigma^{(1)}$ satisfies LME. Analogously as before we arrive at the series
of conserved charges $\sigma^{(1)}, \sigma^{(2)},...$ and currents 
$J^{(1)}, J^{(2)},...$ which are defined by the following recursion equations
\bea
\label{recursion}
J^{(n+1)}_{x}&=&-\partial_{y}\sigma^{(n+1)}={\cal L}_{\tilde{x}}\sigma^{(n)}\nonumber\\
J^{(n+1)}_{y}&=&\partial_{x}\sigma^{(n+1)}={\cal L}_{\tilde{y}}\sigma^{(n)}\h ,\h\h
n=0,1,...
\eea
From the assumption that $\Theta$ is a solution of ME and $\sigma^{(0)}$ satisfies LME
(\ref{LME}) it follows that all $\sigma^{(n)}$s satisfy LME
\be
\partial_{x}\partial_{\tilde{x}}\sigma^{(n)}+
\partial_{y}\partial_{\tilde{y}}\sigma^{(n)}+
\{\partial_{x}\Theta,\partial_{y}\sigma^{(n)}\}_{{\cal M}}-
\{\partial_{y}\Theta,\partial_{x}\sigma^{(n)}\}_{{\cal M}}=0\h ,\h\h n=0,1,...
\ee
This means that $\sigma^{(n)}$, $n=0,1,...$ are the \it hidden symmetries \rm of ME.

\noindent Equation (\ref{recursion}) defines the \it recursion operator \rm $R$ by
\be
\sigma^{(n+1)}=R\,\sigma^{(n)}
\ee
(Compare with \cite{11}).

\noindent For example taking
\be
\label{2.24}
\sigma^{(0)}=\tilde{x}
\ee
we find
\bea
\label{2.25}
\sigma^{(1)}&=&-y+f^{(1)}(\hbar;\tilde{x},\tilde{y},p,q),\nonumber\\
\sigma^{(2)}&=&x\,\partial_{\tilde{y}}f^{(1)}-y\,\partial_{\tilde{x}}f^{(1)}+
\{\Theta,f^{(1)}\}_{{\cal M}}+f^{(2)}(\hbar;\tilde{x},\tilde{y},p,q),\nonumber\\
...etc& &
\eea
\noindent If one puts
\be
\label{2.26}
\sigma^{(0)}=\tilde{y}
\ee
then
\bea
\label{2.27}
\sigma^{(1)}&=&x+f^{(1)}(\hbar;\tilde{x},\tilde{y},p,q),\nonumber\\
\sigma^{(2)}&=&x\,\partial_{\tilde{y}}f^{(1)}-y\,\partial_{\tilde{x}}f^{(1)}+
\{\Theta,f^{(1)}\}_{{\cal M}}+f^{(2)}(\hbar;\tilde{x},\tilde{y},p,q),\nonumber\\
...etc& &
\eea
%
%
\section{Linear systems for ME and dressing operators.}
\setcounter{equation}{0}
We deal here with conserved charges $\eta^{(0)},\eta^{(1)},...,$etc., defined by 
(\ref{eta-zero=1}), (\ref{eta-1-theta}) and (\ref{eta-integralform}). For this especial
choice we put instead of $\eta^{(0)},\eta^{(1)},...,\eta^{(n)},...$ the symbols
$\psi^{(0)},\psi^{(1)},...,\psi^{(n)},...$ so we have
\bea
\psi^{(0)}&=&1 \h ,\h\h\h\h\h\ \psi^{(1)}=\frac{1}{i\hbar}\Theta\h ,\h\h\h\h\h\h\nonumber\\
\psi^{(n)}&=&\int^{x}dx^{(n)}{\cal D}_{\tilde{y}}\int^{x^{(n)}}dx^{(n-1)}{\cal D}_{\tilde{y}}
...\int^{x^{(2)}}dx^{(1)}{\cal D}_{\tilde{y}}1
\eea
Define
\be
\Psi(\lambda):=1+\sum_{n=1}^{\infty}\lambda^{n}\psi^{(n)}\h ,\h\h
\lambda\,\epsilon \overline{\mbox{\boldmath $C$}}-\{\infty\}
\ee
One can easily check that $\Psi(\lambda)$ satisfies the following linear system of
differential equations
\bea
\label{linearsystemforpsi}
\partial_{x}\Psi(\lambda)&=&\lambda{\cal D}_{\tilde{y}}\Psi(\lambda)\nonumber\\
-\partial_{y}\Psi(\lambda)&=&\lambda{\cal D}_{\tilde{x}}\Psi(\lambda)\h ,\h\h
\lambda\,\epsilon \overline{\mbox{\boldmath $C$}}-\{\infty\}
\eea
and, in fact, this system is a Lax pair for ME. Emploing the results of \cite{12,13,14}
one can show that $\Psi(\lambda)$ has the form of
\be
\label{3.4}
\Psi(\lambda)=\exp_{\ast} \{\frac{1}{i\hbar}\sum_{n=1}^{\infty}\lambda^{n}\Lambda^{(n)}\}, \h\h\h\h\h
\Lambda^{(1)}=\Theta 
\ee
\bea
\partial_{x}\Lambda^{(n)}=\partial_{\tilde{y}}\Lambda^{(n-1)}+\sum_{l=1}^{n-1}
\frac{B_{l}}{l!}\sum_{\stackrel{k_{1}+...+k_{l}=n-1}{k_{1},...,k_{l}\geq 1}}
\{\Lambda^{(k_{1})},...,\{\Lambda^{(k_{l})},\partial_{x}\Theta\}_{{\cal M}}...
\}_{{\cal M}} \h\h\h\h \nonumber\\
\partial_{y}\Lambda^{(n)}=-\partial_{\tilde{x}}\Lambda^{(n-1)}+\sum_{l=1}^{n-1}
\frac{B_{l}}{l!}\sum_{\stackrel{k_{1}+...+k_{l}=n-1}{k_{1},...,k_{l}\geq 1}}
\{\Lambda^{(k_{1})},...,\{\Lambda^{(k_{l})},\partial_{y}\Theta\}_{{\cal M}}...
\}_{{\cal M}}, \h n>1 \nonumber
\eea
where $B_{l}$ are the Bernoulli numbers, 
$\frac{t}{e^{t}-1}=\sum_{l=0}^{\infty}B_{l}\frac{t^{l}}{l!}$. The formula (\ref{3.4}) proves
 that if $\Theta$ is analytic in $\hbar$ then all $\Lambda^{(n)}$s are also analytic 
in $\hbar$ \cite{15}.
Then $\Psi_{\ast}^{-1}(\lambda)$ defined by
\be
\Psi_{\ast}^{-1}(\lambda)\ast\Psi(\lambda)=\Psi(\lambda)\ast\Psi_{\ast}^{-1}(\lambda)=1
\ee
has the following form
\be
\Psi_{\ast}^{-1}(\lambda)=\exp_{\ast}
\{-\;\frac{1}{i\hbar}\sum_{n=1}^{\infty}\lambda^{n}\Lambda^{(n)}\}
\ee
and it fulfills the following system
\bea
\label{linearsystemforreverspsi}
\partial_{x}\Psi_{\ast}^{-1}(\lambda)&=&
\lambda(\partial_{\tilde{y}}\Psi_{\ast}^{-1}(\lambda)
-\frac{1}{i\hbar}\Psi_{\ast}^{-1}(\lambda)\ast\partial_{x}\Theta)\nonumber\\
-\partial_{y}\Psi_{\ast}^{-1}(\lambda)&=&
\lambda(\partial_{\tilde{x}}\Psi_{\ast}^{-1}(\lambda)
+\frac{1}{i\hbar}\Psi_{\ast}^{-1}(\lambda)\ast\partial_{y}\Theta)\h ,\h\h
\lambda\,\epsilon \overline{\mbox{\boldmath $C$}}-\{\infty\}
\eea

\vspace{1 ex}

\noindent Analogously, defining
\be
\sigma(\lambda):=\sum_{n=1}^{\infty}\lambda^{n}\sigma^{(n)}\h ,\h\h
\lambda\,\epsilon \overline{\mbox{\boldmath $C$}}-\{\infty\}
\ee
where $\sigma^{(n)}$, $n=0,1,...$ are the conserved charges introduced in the previous 
section (see (\ref{recursion})) one arrives at the system
\bea
\label{3.9}
\partial_{x}\sigma(\lambda)&=&\lambda\; {\cal L}_{\tilde{y}}\sigma(\lambda)\nonumber\\
-\partial_{y}\sigma(\lambda)&=&\lambda\; {\cal L}_{\tilde{x}}\sigma(\lambda)\h ,\h\h
\lambda\,\epsilon \overline{\mbox{\boldmath $C$}}-\{\infty\}
\eea
which is also a Lax pair for ME.

\vspace{1 ex}

\noindent Let $F$ be a function such that
\be
\label{definitionofF}
\sigma(\lambda)=\Psi(\lambda)\ast F\ast\Psi^{-1}_{\ast}(\lambda)
\ee
It is evident that such a function F always exists and is uniquely defined by 
$\sigma(\lambda)$. Moreover, from (\ref{linearsystemforpsi}), 
(\ref{linearsystemforreverspsi}) and (\ref{3.9}) one quickly finds that $F$ must be of the 
form
\be
F=F(\hbar;\tilde{y}+\lambda x , \tilde{x}-\lambda y , \lambda , p , q)
\ee
It means that $F$ is a \it twistor function\rm , as the equations
\be
\tilde{y}+\lambda x =:w^{1}=\mbox{const}\h ,\h\h
\tilde{x}-\lambda y=:w^{2}=\mbox{const}\h ,\h\h
\lambda=:w^{3}=\mbox{const}
\ee
define a totally null anti-self-dual 2-surface in $\mbox{\boldmath $C$}^{4}$ (the \it twistor 
surface\rm ). This twistor surface is the integral manifold for the following 
anti-self-dual 2-form $\omega$
\bea
\omega & = & (d\tilde{y}+\lambda dx)\wedge(d\tilde{x}-\lambda dy)\nonumber\\
 & = & -d\tilde{x}\wedge d\tilde{y}\, + \,\lambda(dx\wedge d\tilde{x} + dy\wedge d\tilde{y})
\, - \,\lambda^{2}dx\wedge dy 
\eea

\vspace{1 ex}

\noindent The formula (\ref{definitionofF}) says that $\Psi(\lambda)$ is the \it dressing
operator \rm for the linear system (\ref{3.9}). 
For example, taking
\be
F:=\tilde{x}-\lambda(y+f^{(1)}(\hbar;p,q))
\ee
one recovers the solution given by (\ref{2.24}), (\ref{2.25}) with 
\mbox{$f^{(1)}=f^{(1)}(\hbar;p,q)$}; taking
\be
F:=\tilde{y}+\lambda(x+f^{(1)}(\hbar;p,q))
\ee
we arrive at (\ref{2.26}), (\ref{2.27}). (Compare with \cite{15}.)

\noindent Analogously as in  our previous work \cite{6} consider the linear system
\bea
\label{3.16}
\frac{1}{\lambda}\partial_{x}\Phi(\frac{1}{\lambda}) & = & 
{\cal D}_{\tilde{y}}\Phi(\frac{1}{\lambda})\nonumber\\
-\frac{1}{\lambda}\partial_{y}\Phi(\frac{1}{\lambda}) & = & 
{\cal D}_{\tilde{x}}\Phi(\frac{1}{\lambda})\h ,\h\h
\lambda\,\epsilon \overline{\mbox{\boldmath $C$}}-\{0\}\\
\Phi(\frac{1}{\lambda}) & = & \Phi^{(0)}+\sum_{n=1}^{\infty}(\frac{1}{\lambda})^{n}
\Phi^{(n)}. \nonumber
\eea
This is also a linear system for ME. One quickly finds that
\bea
\partial_{x}\Theta=i\hbar\; \Phi^{(0)}\ast\partial_{\tilde{y}}[\Phi_{\ast}^{(0)}]^{-1}
\nonumber\\
\partial_{y}\Theta=-i\hbar\; \Phi^{(0)}\ast\partial_{\tilde{x}}[\Phi_{\ast}^{(0)}]^{-1}
\eea
The solution $\Phi(\frac{1}{\lambda})$ can be written in the form
\bea
\Phi(\frac{1}{\lambda})=\exp_{\ast}\{\frac{1}{i\hbar}\sum_{n=0}^{\infty}
(\frac{1}{\lambda})^{n}\Omega^{(n)}\}\nonumber\\
\exp_{\ast}\{\frac{1}{i\hbar}\Omega^{(0)}\}=\Phi^{(0)}.
\eea
Then we consider 
$\tilde{\sigma}(\frac{1}{\lambda})=\sum_{n=0}^{\infty}(\frac{1}{\lambda})^{n}\sigma^{(n)}$
\be
\label{3.19}
\tilde{\sigma}(\frac{1}{\lambda})=\Phi(\frac{1}{\lambda})\ast\tilde{F}\ast
\Phi_{\ast}^{-1}(\frac{1}{\lambda}).
\ee
Straightforward calculations show that $\tilde{\sigma}(\frac{1}{\lambda})$ satisfies the 
following linear system
\bea
\label{3.20}
\frac{1}{\lambda}\partial_{x}\tilde{\sigma}(\frac{1}{\lambda}) & = & 
{\cal L}_{\tilde{y}}\tilde{\sigma}(\frac{1}{\lambda})\nonumber\\
-\frac{1}{\lambda}\partial_{y}\tilde{\sigma}(\frac{1}{\lambda}) & = & 
{\cal L}_{\tilde{x}}\tilde{\sigma}(\frac{1}{\lambda})\h ,\h\h
\lambda\,\epsilon \overline{\mbox{\boldmath $C$}}-\{ 0 \}
\eea
iff the function $\tilde{F}$ is of the form
\be
\tilde{F}=\tilde{F}(\hbar; x+\frac{1}{\lambda}\tilde{y} , -y+\frac{1}{\lambda}\tilde{x} , 
\frac{1}{\lambda} , p , q )
\ee
i.e., $\tilde{F}$ is a twistor function.

\noindent Of course the system (\ref{3.20}) is also a linear system for ME (a Lax pair 
for ME) and Eq. (\ref{3.19}) expresses the fact that $\Phi(\frac{1}{\lambda})$ is the 
dressing operator for this system.
%
%
\section{Infinite algebra of hidden symmetries.}
\setcounter{equation}{0}
From the previous sections one can conclude that the general solution of LME (\ref{LME})
i.e., the general symmetry of ME is given by
\bea
\label{gensymmofsolutME}
\delta_{(F\tilde{F})}\Theta & = & \frac{1}{2\pi i}\oint_{\gamma}
\frac{d\lambda}{\lambda^{2}}
(-\Psi(\lambda)\ast F\ast\Psi_{\ast}^{-1}(\lambda)\, + \, 
\Phi(\lambda)\ast \tilde{F}\ast\Phi_{\ast}^{-1}(\lambda))\nonumber\\
F & = & F(\hbar;\tilde{y}+\lambda x , \tilde{x}-\lambda y , \lambda , p , q)\nonumber\\
\tilde{F} & = & \tilde{F}(\hbar; x+\frac{1}{\lambda}\tilde{y} , 
-y+\frac{1}{\lambda}\tilde{x} , \frac{1}{\lambda} , p , q )
\eea
where a curve $\gamma$ does not contain singularities of functions which are integrated.

\noindent One can compare (\ref{gensymmofsolutME}) with respective formulas given by 
Q.H.Park \cite{2,16}. In (\ref{gensymmofsolutME}) the first sign $(-)$ and the factor 
$\frac{1}{\lambda^{2}}$ are chosen for further convenience.

\vspace{1 ex}

\noindent Now we are looking for the algebra of hidden symmetries of ME. To this end 
consider the commutator
\be
\noindent [\, \delta_{(F_{1}\tilde{F_{1}})}\, ,\, \delta_{(F_{2}\tilde{F_{2}})} ] \Theta =
\ee
\bea 
\delta_{(F_{1}\tilde{F_{1}})}(\Theta+\delta_{(F_{2}\tilde{F_{2}})}\Theta)\; -\;
\delta_{(F_{1}\tilde{F_{1}})}\Theta\; -\;
\delta_{(F_{2}\tilde{F_{2}})}(\Theta+\delta_{(F_{1}\tilde{F_{1}})}\Theta)\; +\;
\delta_{(F_{2}\tilde{F_{2}})}\Theta \nonumber
\eea
Simple calculations with the use of (\ref{gensymmofsolutME}) give
\bea
\label{4.3}
[\, \delta_{(F_{1}\tilde{F_{1}})}\, ,\, \delta_{(F_{2}\tilde{F_{2}})} ] \Theta &=&
\frac{(i\hbar)}{2\pi i}\oint_{\gamma}
\frac{d\lambda}{\lambda^{2}}\,
[\, -\, \{\delta_{(F_{2}\tilde{F_{2}})}\Psi\ast\Psi_{\ast}^{-1}\, , \, 
\Psi\ast F_{1}\ast\Psi_{\ast}^{-1}\}_{{\cal M}}\nonumber\\
& &+\{\delta_{(F_{1}\tilde{F_{1}})}\Psi\ast\Psi_{\ast}^{-1}\, , \, 
\Psi\ast F_{2}\ast\Psi_{\ast}^{-1}\}_{{\cal M}}\nonumber\\
& &+\{\delta_{(F_{2}\tilde{F_{2}})}\Phi\ast\Phi_{\ast}^{-1}\, , \, 
\Phi\ast\tilde{F_{1}}\ast\Phi_{\ast}^{-1}\}_{{\cal M}}\nonumber\\
& &-\{\delta_{(F_{1}\tilde{F_{1}})}\Phi\ast\Phi_{\ast}^{-1}\, , \, 
\Phi\ast\tilde{F_{2}}\ast\Phi_{\ast}^{-1}\}_{{\cal M}}\, ].
\eea
Performing variation of the system (\ref{linearsystemforpsi}) and employing 
(\ref{linearsystemforreverspsi}) one obtains
\bea
\label{4.4}
\partial_{x}(\delta_{(F_{1}\tilde{F_{1}})}\Psi\ast\Psi_{\ast}^{-1})&=&
\lambda\, [\, {\cal L}_{\tilde{y}}(\delta_{(F_{1}\tilde{F_{1}})}\Psi\ast\Psi_{\ast}^{-1})
+\frac{1}{i\hbar}\partial_{x}\delta_{(F_{1}\tilde{F_{1}})}\Theta\, ]\nonumber\\
-\partial_{y}(\delta_{(F_{1}\tilde{F_{1}})}\Psi\ast\Psi_{\ast}^{-1})&=&
\lambda\, [\, {\cal L}_{\tilde{x}}(\delta_{(F_{1}\tilde{F_{1}})}\Psi\ast\Psi_{\ast}^{-1})
-\frac{1}{i\hbar}\partial_{y}\delta_{(F_{1}\tilde{F_{1}})}\Theta\, ]
\eea
The solution of (\ref{4.4}) can be found and it reads
\bea
\label{4.5}
{i\hbar}\delta_{(F_{1}\tilde{F_{1}})}\Psi(\lambda)\ast\Psi_{\ast}^{-1}(\lambda)=
\frac{1}{2\pi i}\sum_{n=1}^{\infty} {\lambda^n} \oint_{\gamma}
\frac{d\lambda'}{(\lambda')^{n+1}}
(-\Psi'\ast F'_1 \ast\Psi_{\ast}^{'-1}\, + \, 
\Phi'\ast \tilde{F'_1}\ast\Phi_{\ast}^{'-1})\nonumber\\
=\frac{1}{2\pi i}\oint_{\gamma'_{>}}d\lambda'\frac{\lambda}{\lambda'(\lambda'-\lambda)}
(-\Psi'\ast F'_1 \ast\Psi_{\ast}^{'-1}\, + \, 
\Phi'\ast \tilde{F'_1}\ast\Phi_{\ast}^{'-1})
\eea
where $\gamma'_{>}$ is any curve closing a domain containing a circle 
\mbox{$\overline{K}(0;r) : \gamma\subset K(0;r)$.} Then $\Psi':=\Psi(\lambda')$, 
\mbox{$F'_{1}:=F_{1}(\hbar;\tilde{y}+\lambda' x , \tilde{x} - \lambda' y , \lambda' ,p,q)$}
,... etc.

Analogously for $\delta_{(F_{1}\tilde{F_{1}})}\Phi\ast\Phi_{\ast}^{-1}$ one gets the
system of equations
\bea
\label{4.6}
\frac{1}{\lambda}\partial_{x}(\delta_{(F_{1}\tilde{F_{1}})}\Phi\ast\Phi_{\ast}^{-1})&=&
 {\cal L}_{\tilde{y}}(\delta_{(F_{1}\tilde{F_{1}})}\Phi\ast\Phi_{\ast}^{-1})
+\frac{1}{i\hbar}\partial_{x}\delta_{(F_{1}\tilde{F_{1}})}\Theta\, \nonumber\\
-\frac{1}{\lambda}\partial_{y}(\delta_{(F_{1}\tilde{F_{1}})}\Phi\ast\Phi_{\ast}^{-1})&=&
 {\cal L}_{\tilde{x}}(\delta_{(F_{1}\tilde{F_{1}})}\Phi\ast\Phi_{\ast}^{-1})
-\frac{1}{i\hbar}\partial_{y}\delta_{(F_{1}\tilde{F_{1}})}\Theta\, 
\eea
The solution of (\ref{4.6}) reads
\be
\label{4.7}
{i\hbar}\delta_{(F_{1}\tilde{F_{1}})}\Phi(\frac{1}{\lambda})\ast\Phi_{\ast}^{-1}
(\frac{1}{\lambda})=
-\frac{1}{2\pi i}\sum_{n=0}^{\infty}\frac{1}{\lambda^{n}}\oint_{\gamma}
d\lambda'(\lambda')^{n-1}(-\Psi'\ast F'_1 \ast\Psi_{\ast}^{'-1}\, + \, 
\Phi'\ast \tilde{F'_1}\ast\Phi_{\ast}^{'-1})
\ee
Since for calculating (\ref{4.3}) we need $\lambda\,\epsilon\gamma$ one can write down
(\ref{4.7}) in the form of
\be
\label{4.8}
{i\hbar}\delta_{(F_{1}\tilde{F_{1}})}\Phi(\frac{1}{\lambda})\ast\Phi_{\ast}^{-1}
(\frac{1}{\lambda})|_{\lambda\,\epsilon\gamma}=
\frac{1}{2\pi i}\oint_{\gamma'_{<}}
d\lambda'\frac{\lambda}{\lambda'(\lambda'-\lambda)}(-\Psi'\ast F'_1 \ast\Psi_{\ast}^{'-1}\, 
+ \, \Phi'\ast \tilde{F'_1}\ast\Phi_{\ast}^{'-1})
\ee
where now $\gamma'_{<}\subset K(0;r)$ and $\overline{K}(0;r)$ is a circle belonging to the
domain closed by $\gamma$. Note that $\gamma$, $\gamma'_{>}$ and $\gamma'_{<}$ must be 
chosen so that they close the same singularities of the integrated functions.
Substituting (\ref{4.5}) and (\ref{4.8}) and also analogous formulas for 
\mbox{$\delta_{(F_{2}\tilde{F_{2}})}\Psi(\lambda)\ast\Psi_{\ast}^{-1}(\lambda)$} and
\mbox{$\delta_{(F_{2}\tilde{F_{2}})}\Phi(\frac{1}{\lambda})\ast\Phi_{\ast}^{-1}
(\frac{1}{\lambda})$} into (\ref{4.3}), performing then integrations and applying the
residue theorem one gets
\be
\label{algebraofhiddensymmetries}
[\, \delta_{(F_{1}\tilde{F_{1}})}\, ,\, \delta_{(F_{2}\tilde{F_{2}})} ] \Theta\, =\,
2 \delta_{(\{F_{1},F_{2}\}_{{\cal M}}\{\tilde{F_{1}},\tilde{F_{2}}\}_{{\cal M}})}
\Theta
\ee
Hence, the \it hidden symmetries for ME form a closed algebra associated with the Moyal
bracket Lie algebra\rm .

\vspace{1 ex}
\noindent This coincides with the results of \cite{2,11,16} where the algebra of hidden
symmetries is given for Pleba\'{n}ski's heavenly equations and for SDYM equations.

As ME can be reduced to all known heavenly equations, to $su(N)$ SDYM equations, to
$su(N)$ chiral equations etc., the algebra (\ref{algebraofhiddensymmetries}) contains
the hidden symmetry algebras for the reduced equations.
%
%
\section{Twistor construction.}
\setcounter{equation}{0}
For completeness we describe here briefly a twistor construction for ME given in our previous
work \cite{6}. This construction is similar to the one for SDYM equations \cite{17,18}.

\noindent What must be noted, and it has been pointed out by K.Takasaki \cite{15}, is 
that the twistor construction we propose is valid for the case of $\Theta$ being analytic 
in $\hbar$, i.e., \mbox{$\Theta=\sum_{n=0}^{\infty}\hbar^{n}\Theta_{n}$}, 
$\Theta_{n}=\Theta_{n}(x,y,\tilde{x},\tilde{y},p,q)$. The case with negative powers of 
$\hbar$ should be considered separately and as we know from \cite{6} it can be solved by 
the Fairlie-Leznov method \cite{19} but we still have not succeeded in a twistor image for this 
case.

We start with a twistor function
\bea
\label{5.1}
&H=H(\lambda)=H(\hbar; \tilde{y}+\lambda x,\tilde{x}-\lambda y,\lambda,p,q)
=\exp_{\ast}\{\frac{1}{i\hbar}\sum_{n=-\infty}^{\infty}
\lambda^{n}\Delta^{(n)}\}&\nonumber\\
&\lambda\,\epsilon\,(\overline{\mbox{\boldmath $C$}}-\{0\})\cap 
(\overline{\mbox{\boldmath $C$}}-\{\infty\})&\nonumber\\
&\Delta^{(n)}=\sum_{m=0}^{\infty}\hbar^{m}\Delta^{(n)}_{m}(x,y,\tilde{x},\tilde{y},p,q) &
\eea
Let $\Psi=\Psi(\lambda)$ be a function of the form
\bea
\label{gauge-5.2}
&\Psi(\lambda)=\exp_{\ast}\{\frac{1}{i\hbar}\sum_{n=1}^{\infty}
\lambda^{n}\Lambda^{(n)}\}\, ,\h\h\h\h\lambda\,\epsilon\,(\overline{\mbox{\boldmath $C$}}-\{\infty\})&\nonumber\\
&\Lambda^{(n)}=\Lambda^{(n)}(\hbar;x,y,\tilde{x},\tilde{y},p,q)=
\sum_{m=0}^{\infty}\hbar^{m}\Lambda^{(n)}_{m}(x,y,\tilde{x},\tilde{y},p,q)\, ,&\nonumber\\
&\Lambda^{(1)}=\Theta &
\eea
and let $\Phi=\Phi(\frac{1}{\lambda})$ be a function of the form
\bea
&\Phi(\frac{1}{\lambda})=\exp_{\ast}\{\frac{1}{i\hbar}\sum_{n=1}^{\infty}
(\frac{1}{\lambda})^{n}\Omega^{(n)}\}\, ,\h\h\h\h
\lambda\,\epsilon\,(\overline{\mbox{\boldmath $C$}}-\{0\})&\nonumber\\
&\Omega^{(n)}=\Omega^{(n)}(\hbar;x,y,\tilde{x},\tilde{y},p,q)=
\sum_{m=0}^{\infty}\hbar^{m}\Omega^{(n)}_{m}(x,y,\tilde{x},\tilde{y},p,q)\, ,&
\eea
These functions are chosen so that the following factorization holds
\be
H(\lambda)=\Phi^{-1}_{\ast}(\frac{1}{\lambda})\ast\Psi(\lambda)
\h, \h\h\h
\lambda\,\epsilon\,(\overline{\mbox{\boldmath $C$}}
-\{0\})\cap (\overline{\mbox{\boldmath $C$}}-\{\infty\}).
\ee
(This is the \it Riemann-Hilbert problem \rm or the \it Birkhoff factorization \rm \cite{18}).

\noindent One easily finds that from the conditions:
\mbox{$(\lambda\partial_{\tilde{y}}-\partial_{x})H(\lambda)=0$} and
\mbox{$(\lambda\partial_{\tilde{x}}+\partial_y)H(\lambda)=0$}
it follows that 
\bea
& [(\lambda\partial_{\tilde{y}}-\partial_{x})\Psi(\lambda)]
\ast\Psi^{-1}_{\ast}(\lambda)=\lambda[(\partial_{\tilde{y}}
-\frac{1}{\lambda}\partial_x )\Phi(\frac{1}{\lambda})]\ast
\Phi^{-1}_{\ast}(\frac{1}{\lambda}) &\nonumber\\
& [(\lambda\partial_{\tilde{x}}+\partial_{y})\Psi(\lambda)]
\ast\Psi^{-1}_{\ast}(\lambda)=\lambda[(\partial_{\tilde{x}}
+\frac{1}{\lambda}\partial_y )\Phi(\frac{1}{\lambda})]\ast
\Phi^{-1}_{\ast}(\frac{1}{\lambda}) & \nonumber\\
& \lambda\,\epsilon\,(\overline{\mbox{\boldmath $C$}}-\{0\})\cap
(\overline{\mbox{\boldmath $C$}}-\{\infty\}) &
\label{5.5}
\eea
The left-hand side of Eq.~(\ref{5.5}) can be analytically extended on 
$\overline{\mbox{\boldmath $C$}}$ and in the 
gauge (\ref{gauge-5.2}) we get
\bea
& [(\lambda\partial_{\tilde{y}}-\partial_{x})\Psi(\lambda)]
\ast\Psi^{-1}_{\ast}(\lambda)=-\lambda \frac{1}{i\hbar}\partial_{x}\Theta & \nonumber\\
& [(\lambda\partial_{\tilde{x}}+\partial_{y})\Psi(\lambda)]
\ast\Psi^{-1}_{\ast}(\lambda)=\lambda \frac{1}{i\hbar}\partial_{y}\Theta \, ,
\hspace{.5 em} \lambda\,\epsilon\,\overline{\mbox{\boldmath $C$}}&
\label{5.6}
\eea
Thus we recover the linear system (\ref{linearsystemforpsi}) of ME. Substituting 
(\ref{5.6}) into (\ref{5.5}) one recovers the linear system (\ref{3.16}) of ME as well.

It means that our procedure gives a twistor construction for ME
\section*{Acknowledgments}
We are grateful to Maciej Dunajski for many discussions on the problems considered in 
the paper. 

\noindent This work was partially supported by the CONACyT (M\'{e}xico) grant 32427-E
and by KBN (Poland) grant Z/370/S. 

\end{document}